\documentclass[aps,twocolumn,floatfix,superscriptaddress,showpacs,preprintnum
bers,amsmath,amssymb,showpacs,showkeys]{revtex4}
\usepackage{amssymb}
\usepackage{amsmath}
\usepackage{amsfonts}
\usepackage{epsfig}
\usepackage{graphicx}
\usepackage{tabularx}
\newcommand{\seq}{\begin{subequations}}
\newcommand{\sen}{\end{subequatons}}
\newcommand{\eq}{\begin{eqnarray}}
\newcommand{\en}{\end{eqnarray}}

\def\shiftdown#1{#1\llap{\lower.04ex\hbox{#1}}}

\newcommand{\bfrI}{{\bf r}}

\newcommand{\bfkI}{{\bf k}}
\newcommand{\bfksI}{{\bf k'}}

\newcommand{\bfk}{{\bf k}_{\perp}}
\newcommand{\bfkpr}{{\bf k}_{\perp}^\prime}

\newcommand{\bfb}{{\bf b}_{\perp}}

\begin{document}

\title{Light-front potential for heavy quarkonia\\ 
constrained by the holographic soft-wall model} 

\author{Thomas Gutsche}
\affiliation{Institut f\"ur Theoretische Physik,
Universit\"at T\"ubingen, \\
Kepler Center for Astro and Particle Physics, \\ 
Auf der Morgenstelle 14, D-72076 T\"ubingen, Germany}
\author{Valery E. Lyubovitskij} 
\affiliation{Institut f\"ur Theoretische Physik,
Universit\"at T\"ubingen, \\
Kepler Center for Astro and Particle Physics, \\
Auf der Morgenstelle 14, D-72076 T\"ubingen, Germany}
\affiliation{Department of Physics, Tomsk State University,  
634050 Tomsk, Russia} 
\affiliation{Mathematical Physics Department, 
Tomsk Polytechnic University, \\
Lenin Avenue 30, 634050 Tomsk, Russia} 
\author{Ivan Schmidt}
\affiliation{Departamento de F\'\i sica y Centro Cient\'\i
fico Tecnol\'ogico de Valpara\'\i so (CCTVal), Universidad T\'ecnica
Federico Santa Mar\'\i a, Casilla 110-V, Valpara\'\i so, Chile}
\author{Alfredo Vega}
\affiliation{Instituto de F\'isica y Astronom\'ia, 
Universidad de Valpara\'iso, \\
Avenida Gran Breta\~na 1111, Valpara\'iso, Chile}
\affiliation{Centro de Astrof\'isica de Valpara\'iso, \\
Avenida Gran Breta\~na 1111, Valpara\'iso, Chile} 

\vspace*{.2cm}

\date{\today}

\begin{abstract}

We derive a light-front Schr\"odinger-type equation of motion 
for the quark-antiquark wave function of heavy quarkonia 
imposing constraints from the holographic soft-wall model. 

\end{abstract}



\maketitle

\section{Introduction}

The main objective of this paper is to derive a Schr\"odinger 
type equation of motion (EOM) for the quark-antiquark wave function of 
heavy quarkonia imposing constraints from the holographic soft-wall model. 
Recently this topic received additional interest because of the discovery of
gauge-gravity duality, which helps to establish an equivalence between 
gravity on  anti-de Sitter (AdS) space and light-front QCD -- light-front 
holography~\cite{deTeramond:2008ht}. Using light-front 
holography one can derive light-front wave functions for light 
and heavy hadrons~\cite{Brodsky:2007hb}-\cite{Gutsche:2013zia}. 
The light-front wave functions obey a Schr\"odinger-type EOM
with an effective confinement potential which is quadratically dependent 
on the transverse coordinate. This EOM produces 
linear Regge-trajectories for hadronic masses. 
Recently, in Ref.~\cite{Trawinski:2014msa}, it was shown that this 
quadratic potential is related to an instant-form (IF) 
effective potential (after squaring). It is consistent with effective 
potentials obtained in lattice gauge theory and from string models of 
hadrons, including an approximate agreement with the numerical value 
of the strength of these potentials of $\simeq$ 400 MeV. Another interesting 
application of the gauge/gravity duality to heavy-quark potentials  
has been pursued in Ref.~\cite{Andreev:2006ct}. 

Here we would like to continue our analysis of the 
light-front (LF) Schr\"odinger-type equation 
for heavy quarkonia started in 
Refs.~\cite{Vega:2009zb}-\cite{Gutsche:2012ez}. In particular, 
in these papers we derived 
a LF Schr\"odinger-type equation for mesons, containing 
light and heavy quarks, from a holographic soft-wall 
model~\cite{Vega:2009zb}-\cite{Gutsche:2012ez},%
\cite{Vega:2010ns}-\cite{Gutsche:2012bp}. The corresponding 
wave functions produce the correct Regge-behavior of  hadronic mass specta 
and are consistent with constraints imposed by chiral symmetry in 
the light quark sector and heavy quark symmetry in the heavy 
quark sector. In particular, for light pseudoscalar mesons the masses 
satisfy the Gell-Mann-Oakes-Renner and Gell-Mann-Okubo relations. 
In the sector of heavy quarks we get agreement with heavy quark effective 
theory and potential models for heavy quarkonia.  
In the heavy quark mass limit 
$m_Q \to \infty$ we obtain the correct scaling of 
leptonic decay constants both for heavy-light mesons 
$f_{Q\bar q} \sim 1/\sqrt{m_Q}$ and for heavy quarkonia  
$f_{Q\bar Q} \sim \sqrt{m_Q}$ and 
$f_{c\bar b} \sim m_c/\sqrt{m_b}$ with $m_c \ll m_b$. 
In this limit we also generate the correct expansion of heavy meson masses 
\eq\label{mass_expansion} 
M_{Q\bar q} &=& m_Q + \bar\Lambda + {\cal O}(1/m_Q) \,, \nonumber\\
M_{Q_1 \bar Q_2} &=& m_{Q_1} + m_{Q_2} + E 
+ {\cal O}(1/m_{Q_{1,2}}) \,, 
\en 
where the quantities $\bar\Lambda$ and $E$ are the ${\cal O}(1)$ 
contributions to the masses of heavy mesons.
Their splittings, e.g. between the vector and 
pseudoscalar states of heavy-light mesons, become  
\eq 
M_{Q\bar q}^V - M_{Q\bar q}^P \, \sim \, \frac{1}{m_Q} \,. 
\en  
Notice that in Refs.~\cite{Vega:2009zb}-\cite{Gutsche:2012ez} the effective 
LF potential consists of two parts: a nonperturbative part  --   
with a confining potential dictated by the holographic soft-wall model, and 
a perturbative part -- the color Coulomb potential. It is important to 
note that the presence of these two parts in the LF potential 
has been confirmed by a recent analysis performed 
in Ref.~\cite{Trawinski:2014msa}. 

In this paper, using our results obtained in 
Refs.~\cite{Vega:2009zb}-\cite{Gutsche:2012ez} and those of 
Ref.~\cite{Trawinski:2014msa}, we focus on the connection between the IF 
potential (the sum of a linearly rising confinement potential and 
the perturbative color Coulomb potential) and the potential in  
LF QCD, used for the evaluation of LF wave functions and 
for predictions of hadronic mass spectra. For simplicity we concentrate 
on the case of heavy quarkonia systems. 

\section{Light-front equation of motion for heavy quarkonia} 

We start from the IF EOM in three-dimensional momentum 
space, describing a bound state of a heavy quark $Q$ and an antiquark $\bar Q$ 
\eq\label{EOM_IF_Operator} 
2 E_k \, \psi(\bfkI) 
+ \int\!\!\frac{d^3\bfksI}{(2\pi)^3 E_{k'}} \, V(\bfkI-\bfksI) 
\, \psi(\bfksI) = M \, \psi(\bfkI) 
\en 
or in matrix form 
\eq\label{EOM_IF_matrix} 
& &\int\!\!\frac{d^3\bfkI}{(2\pi)^3 E_{k}} 
\psi^\dagger(\bfkI) (2 E_k - M) \psi(\bfkI) \nonumber\\
&=& 
- \int\!\!\frac{d^3\bfkI \, d^3\bfksI}{(2\pi)^6 E_{k} E_{k'}}   
\, \psi^\dagger(\bfkI) V(\bfkI-\bfksI) \psi(\bfksI)\, . 
\en 
$M$ and $m_Q$ are the masses of the heavy meson and heavy quark/antiquark, 
respectively, $\psi(\bfkI)$ is the IF wave function and 
$E_k = \sqrt{\bfkI^2 + m_Q^2}$ is the heavy quark energy. 
We work in the rest frame of the heavy quarkonia with
\eq 
{\bf k}_Q &=& - {\bf k}_{\bar Q} \ = \ {\bf k}\,, \nonumber\\
{\bf k}_{\perp, Q} &=& - {\bf k}_{\perp, \bar Q} \ = \ \bfk\,, 
\nonumber\\
k^3_Q &=& - k^3_{\bar Q} \ = \ k^3\,. 
\en  
The IF wave function obeys the normalization condition 
\eq 
\int\!\!\frac{d^3\bfkI}{(2\pi)^3 E_{k}} \, |\psi(\bfkI)|^2 = 1 \,. 
\en 
$V(\bfkI-\bfksI)$ is the effective IF potential in momentum space. 
Its Fourier transform $V(\bfrI)$ in coordinate space 
has a very clear interpretation (see recent discussion 
in Ref.~\cite{Trawinski:2014msa} in terms of the nonperturbative
part of the Cornell potential~\cite{Eichten:1978tg}), which is  
a linearly rising confinement potential confirmed 
by QCD lattice calculations as 
\eq
\frac{V(\bfkI-\bfksI)}{E_k E_{k'}} &=& 
\int\!\! d^3\bfrI e^{i\bfrI (\bfkI-\bfksI)} \, V(\bfrI)\,, \nonumber\\
V(\bfrI) &\equiv& V_{\rm nonpert}(r) = V_0 + \sigma r\, .
\en   
$V_0$ is a constant term and $\sigma$ is a parameter, 
which is identified with the string tension in lattice QCD 
(see e.g. the discussion in Ref.~\cite{Kawanai:2011xb}).  

This equation can be written in the LF frame 
using the relation between IF $\psi(\bfkI)$ and LF 
$\psi(x,\bfk)$ wave functions (see discussion about 
this issue e.g. in Refs.~\cite{Karmanov:1979if,Miller:2009fc})   
\eq 
\psi(\bfkI) \to \sqrt{x (1-x)} \, \psi(x,\bfk) 
\en 
and the expression for the three-momentum component $k^3$ 
in terms of $\bfk$ and the light-cone coordinates~\cite{Jaus:1991cy} 
(see details in Appendix~A.1) 
\eq\label{Jaus_eq} 
k^3 = \frac{x-1/2}{\sqrt{x(1-x)}} \, \sqrt{\bfk^2+m_Q^2}\,. 
\en 
The Jacobian of the variable transformation $k^3 \to x$
is 
\eq
\frac{\partial k^3}{\partial x} = \frac{E_k}{2 x (1-x)} \,, 
\en 
which leads to a change of the integration measure 
\eq
\frac{d^3\bfkI}{E_k} \to \frac{d^2\bfk dx}{2 x (1-x)}\,.
\en 
The effective potential in IF transforms  
into the effective potential in the LC frame as 
\eq
V(\bfkI-\bfksI) \to V(x,\bfk-\bfkpr)
\delta(1-x-x') x (1-x) \,, 
\en 
where the delta-function $\delta(1-x-x')$ imposes 
that $x + x' = 1$ up to $1/m_Q$ corrections 
(see details in Appendix~A.2). 
Therefore, in the LF frame the EOM for heavy quarkonia reads 
\eq 
\hspace*{-1cm}
& &\!\!\!\!\int\limits_0^1\!\! dx \!\!\int\!\! \frac{d^2\bfk}{16 \pi^3} 
\psi^\dagger(x,\bfk) 
\biggl(\sqrt{\frac{\bfk^2+m_Q^2}{x (1-x)}}- M \biggr) 
\psi(x,\bfk) 
\nonumber\\
\hspace*{-1cm}
&=&\!\!\!\!\, - \!\!\int\limits_0^1\!\! dx \!\!
\int\!\! \frac{d^2\bfk \, d^2 \bfkpr}{(16 \pi^3)^2} 
\psi^\dagger(x,\bfk) V(x,\bfk-\bfkpr)\psi(x,\bfkpr) \,.\nonumber\\
& &
\en 
In conjugated two-dimensional impact space the LF EOM reads  
\eq
\hspace*{-.25cm}
& &\!\!\!
\int\limits_0^1\!\! dx \!\! \int\!\! d^2\bfb 
\psi^\dagger(x,\bfb) 
\biggl(\sqrt{\frac{-\partial^2_{\bfb}+m_Q^2}{x (1-x)}}- M \biggr) 
\psi(x,\bfb) 
\nonumber\\
\hspace*{-.25cm}
&=&\!\!\!- \!\int\limits_0^1\!\! dx\!\! \int\!\! d^2\bfb 
\psi^\dagger(x,\bfb) V(x,\bfb)
\psi(x,\bfb) \,. 
\en 
The corresponding normalization conditions for the 
light-front wave functions are given as 
\eq 
& &\!\!\!\int\limits_0^1\!\! dx \!\!\int\!\! \frac{d^2\bfk}{16 \pi^3} 
|\psi(x,\bfk)|^2 = 1  \,,\nonumber\\
& &\!\!\!\int\limits_0^1\!\! dx \!\!\int\!\! d^2\bfb 
|\psi(x,\bfb)|^2 = 1  \,. 
\en 
Next we derive the squared EOMs. For this aim we drop the terms of 
order $\bfk^2/m_Q$ or $\partial^2_{\bfb}/m_Q$, which appear 
in the $1/m_Q$ expansion of the mixed terms 
\eq 
V(x,\bfk-\bfkpr) \sqrt{\bfk^2+m_Q^2} &=& 
V(x,\bfk-\bfkpr)  m_Q \,+\,\ldots\,\nonumber\\
V(x,\bfb)  \sqrt{-\partial^2_{\bfb}+m_Q^2} &=& 
V(x,\bfb)  m_Q \,+\,\ldots 
\en 
Using the counting of kinematical variables in the 
$1/m_Q$ expansion, as discussed in Appendix~A.2,
the squared EOM (up to order $1/m_Q$ corrections) in impact space reads 
\eq\label{m2general}
M^2&=&\int\limits_0^1\!\!dx \!\! \int\!\! d^2\bfb \, 
\psi^\dagger(x,\bfb) \, \hat{\cal M}^2 \, \psi(x,\bfb) \,, 
\nonumber\\
 \hat{\cal M}^2 &=& 
- \frac{\partial^2_{\bfb}}{x (1-x)} 
+\biggl[ \frac{m_Q}{\sqrt{x (1-x)}} + V(x,\bfb) \biggr]^2 
\en 
where $\hat{\cal M}^2$ is the effective LF operator producing 
the mass squared of heavy quarkonia. 
The effective operator $\hat{\cal M}^2$ contains three terms --- 
the invariant mass operator of the two-partonic state $\hat{\cal M}^2_0$
\eq 
\hat{\cal M}^2_0 = \frac{-\partial^2_{\bfb}+m_Q^2}{x (1-x)}, 
\en  
the squared potential 
\eq 
U(x,\bfb) = V^2(x,\bfb)
\en 
and a mixed term 
\eq 
2 V(x,\bfb) \frac{m_Q}{\sqrt{x (1-x)}} \,. 
\en 
In Ref.~\cite{Trawinski:2014msa} it was stressed that
the soft-wall potential corresponds to the 
squared part of the nonperturbative Cornell potential. 
This is true when we consider the massless case and drop 
the constant term $V_0$ in the Cornell potential. 

In the present manuscript we derive an effective hea\-vy quar\-konia 
potential for finite values of the heavy quark masses using 
constraints imposed by soft-wall AdS/QCD and the Cornell potential.   
We remind that the meson mass spectrum was studied in detail 
in the soft-wall AdS/QCD model in~\cite{deTeramond:2008ht,Brodsky:2007hb,%
Branz:2010ub,Gutsche:2012ez}. It was proposed to introduce 
the holographic coordinate $\zeta$ which is related to the 
impact variable $\bfb$ via a holographic 
mapping~\cite{deTeramond:2008ht,Brodsky:2007hb}   
\eq 
\zeta^2 = \bfb^2 x (1-x) \,. 
\en 
Then the effective light-front wave function $\psi(x,\zeta)$, depending 
on the $\zeta$ and $x$ variables is factorized into transverse $\Phi(\zeta)$ 
and longitudinal $f(x,m_Q)$ modes~\cite{deTeramond:2008ht,Brodsky:2007hb,%
Branz:2010ub,Gutsche:2012ez} 
\eq 
& &\psi(x,\bfb) \to \psi(x,\zeta)\,, \\
\vspace*{-1cm}
& &\psi(x,\zeta) \, \sqrt{\frac{2\pi\zeta}{x (1-x)}} \equiv 
\Phi(x,\zeta) = \Phi(\zeta) \, 
f(x,m_Q) \,, \nonumber
\en 
where 
\eq 
\Phi(\zeta) = \kappa^{1+L} \sqrt{\frac{2n!}{(n+L)!}} \, 
\zeta^{1/2+L} \, e^{-\kappa^2\zeta^2/2} \, L_n^L(\kappa^2\zeta^2)\,,  
\en 
and $\kappa$ is the dimensional dilaton parameter. 
In the massless case the longitudinal mode fulfills
$f(x,0) = 1$~\cite{deTeramond:2008ht,Brodsky:2007hb}. 
We studied the massive case $m_Q \neq 0$ in Ref.~\cite{Gutsche:2012ez} and 
showed that the choice of the longitudinal wave function $f(x,m_Q)$ 
can absorb the contribution of the mixing term $2V(\zeta) m_Q/\sqrt{x (1-x)}$
in the effective potential.
At the same time this procedure guarantees 
the agreement of leptonic decay constants and masses of heavy quarkonia 
with the heavy-quark mass expansion (see discussion in Sec.I). 
In particular, the required form for the longitudinal mode $f(x,m_Q)$ is 
\eq 
f(x,m_Q) = N x^\alpha (1-x)^\alpha\,, \quad 
\alpha = \frac{m_Q}{4E} - 1
\en 
where $E$ is the parameter defining the ${\cal O}(m_Q^0)$ contribution to
the heavy quarkonia mass and $N$ is a normalization constant 
fixed by the condition 
\eq 
\int\limits_0^1\!\! dx f^2(x,m_Q) = 1 \,. 
\en   
The equation for the wave function $\Phi(x,\zeta)$ then reads 
\eq\label{V_general} 
M^2&=&\int\limits_0^1\!\! dx \!\! \int\limits_0^\infty\!\! d\zeta \, 
\Phi^\dagger(x,\zeta) \, \hat{\cal M}^2 \, \Phi(x,\zeta)\,, \\
\hat{\cal M}^2 &=& -\partial^2_{\zeta} + \frac{m_Q^2}{x (1-x)} 
+ V^2(\zeta) \,. \nonumber
\en 
The expression for the corresponding operator $\hat{\cal M}^2$ 
in soft-wall AdS/QCD model~\cite{deTeramond:2008ht}-\cite{Branz:2010ub} 
has the form   
\eq\label{V_softwall}  
\hat{\cal M}^2 &=& -\partial^2_{\zeta} + \frac{m_Q^2}{x (1-x)} \nonumber\\
&+& \frac{4L^2-1}{4\zeta^2} + 2\kappa^2 (J-1) + U_{\rm SW}(\zeta) \,. 
\en 
Comparing the two mass operators $\hat{\cal M}^2$, in the general case
of Eq. (\ref{m2general}) and in the soft-wall model (\ref{V_softwall}),
following conclusions can be reached: 
1) both expressions have the same kinetic term; 2) the terms 
$(4L^2-1)/4\zeta^2$ and $ 2\kappa^2 (J-1)$ in the soft-wall approach clearly 
take into account the $L$ and $J$ dependence of the hadronic mass operator; 
3) as was shown in Ref.~\cite{Trawinski:2014msa}, the soft-wall AdS/QCD 
potential $U_{\rm SW}(\zeta)$ corresponds to the square of the 
linearized part of the Cornell potential, $U_{\rm SW}(\zeta)$ has the 
form $U_{\rm SW}(\zeta) = \kappa^4 \zeta^2$~\cite{deTeramond:2008ht,%
Brodsky:2007hb,Vega:2009zb,Branz:2010ub}; 
4) apart from the nonperturbative terms there are important 
corrections generated by the perturbative part of the Cornell 
potential --- the color Coulomb potential 
$V_{\rm pert} = - \frac{4}{3} \alpha_s/r$.  
We included such corrections in our 
previous papers~\cite{Branz:2010ub,Gutsche:2012ez} in 
the form of a constant term~\cite{Sergeenko:1993sn} with  
\eq 
M^2_{\rm pert} = - \frac{64}{9} \frac{\alpha_s^2 m_Q^2}{(n+L+1)^2}\,. 
\en 
These corrections are quite important in order to describe for example
deviations of the mass trajectories of bottom quarkonia from 
the Regge-like ones.  

Now we are in a position to write down the heavy quarkonia mass operator 
squared including constraints imposed both by the soft-wall AdS/QCD model and 
the Cornell potential 
\eq 
\hat{\cal M}^2 &=& 
-\partial^2_{\zeta} + \frac{m_Q^2}{x (1-x)} + 
 \frac{4L^2-1}{4\zeta^2} + 2\kappa^2 (J-1) \nonumber\\
&+& \kappa^4 \zeta^2 
- \frac{64}{9} \frac{\alpha_s^2 m_Q^2}{(n+L+1)^2} \,. 
\en 
The extension to heavy quarks with different flavors $Q_1 \neq Q_2$ 
is straightforward. One should use the longitudinal wave function 
\eq 
\hspace*{-.5cm}
f(x,m_{Q_1},m_{Q_2}) &=& N x^{\alpha_1} (1-x)^{\alpha_2}\,, \nonumber\\
\hspace*{-.5cm}
\alpha_{Q_i} &=& \frac{m_{Q_i}}{4 E} \, 
\biggl( 1 - \frac{E}{2 (m_{Q_1} + m_{Q_2})} \biggr) 
\en  
which reduces to $f(x,m_Q)$ for $m_{Q_1} = m_{Q_2}$. 
The expression for the mass operator squared is modified as 
\eq 
\hat{\cal M}^2 &=& 
-\partial^2_{\zeta} + \frac{m_{Q_1}^2}{x} + \frac{m_{Q_2}^2}{1-x} 
+ \frac{4L^2-1}{4\zeta^2} + 2\kappa^2 (J-1) \nonumber\\
&+& \kappa^4 \zeta^2 
- \frac{64}{9} \frac{\alpha_s^2 m_{Q_1} m_{Q_2}}{(n+L+1)^2} \,. 
\en 
Finally, the mass spectrum for heavy quarkonia is given by 
\eq 
M_{Q_1\bar Q_2}^2 &=& 4 \kappa^2 
\biggl( n + \frac{L + J}{2}\biggr) \nonumber\\
&+&(1 + 2\alpha_1 + 2\alpha_2) 
\biggl(\frac{m_{Q_1}^2}{2\alpha_1} + 
       \frac{m_{Q_2}^2}{2\alpha_2}
\biggr) \nonumber\\
&-& \frac{64}{9} \frac{\alpha_s^2 m_{Q_1} m_{Q_2}}{(n+L+1)^2} \,. 
\en 
This master formula can be further simplified when we drop 
the $O(1/m_Q)$ corrections resulting in 
\eq 
M_{Q_1\bar Q_2}^2 &=& 4 \kappa^2 
\biggl( n + \frac{L + J}{2}\biggr) + (m_{Q_1} + m_{Q_2} + E)^2 
\nonumber\\
&-& \frac{64}{9} \frac{\alpha_s^2 m_{Q_1} m_{Q_2}}{(n+L+1)^2} 
+ {\cal O}(1/m_{Q_{1,2}})
\,. 
\en 
Our results for the mass spectrum of $c\bar c$, $b \bar b$ 
and $c \bar b$ quarkonia are shown in Table I. 
We present our predictions for ground and excited states for 
different values of $n$, $L$ and $J$. 
In the numerical analysis we use the following set of QCD 
and model parameters: 
charm and bottom quark masses 
\eq\label{quark_masses}
m_c = 1.275 \ {\rm GeV}\,, \hspace*{.5cm}
m_b = 4.18  \ {\rm GeV}\,, 
\en 
strong coupling constants 
\eq 
\alpha_s(c\bar c) = 0.45\,, \hspace*{.2cm}
\alpha_s(c\bar b) = 0.383\,, \hspace*{.2cm}
\alpha_s(b\bar b) = 0.27 \,,  
\en 
parameters defining the ${\cal O}(m_Q^0)$ contributions to 
heavy quarkonia masses 
\eq 
& &E_{cc} = 0.795 \ {\rm GeV}\,, \hspace*{.2cm} 
E_{cb} = 1.25 \ {\rm GeV}\,, \nonumber\\ 
& &E_{bb} = 1.45 \ {\rm GeV}\,,
\en 
and the dilaton parameters 
\eq
& &\kappa_{cc} = 0.610\ {\rm GeV}\,, \hspace*{.2cm} 
\kappa_{cb} = 0.629 \ {\rm GeV}\,, \nonumber\\ 
& &\kappa_{bb} = 1.079 \ {\rm GeV}\,. 
\en
Note the masses of the ground states of mesons in Table I are for the fit 
of free parameters, while the results for the excited states are our 
predictions.

\begin{table}
\caption{Masses of heavy quarkonia $c\bar c$, $b\bar b$ and
$c \bar b$}

\label{tab:masses}\def\arraystretch{1.2}
\begin{tabular}{|l|c|c|c|c|l|l|l|l|}
\hline
Meson&$J^{\rm P}$&$n$&$L$&$S$&\multicolumn{4}{c|}{Mass [MeV]} \\ \hline
$\eta_c(2980)$&$0^{-}$ &0,1,2,3&0&0  & 2975 & 3477 & 3729 & 3938 \\\hline
$\psi(3097)$ &$1^{-}$ &0,1,2,3&0&1  & 3097 & 3583 & 3828 & 4032 \\ \hline
$\chi_{c0}(3415)$&$0^{+}$&0,1,2,3&1&1& 3369 & 3628 & 3843 & 4038 \\ \hline
$\chi_{c1}(3510)$&$1^{+}$&0,1,2,3&1&1& 3477 & 3729 & 3938 & 4129 \\ \hline
$\chi_{c2}(3555)$&$2^{+}$&0,1,2,3&1&1& 3583 & 3828 & 4032 & 4219 \\ \hline
$\eta_{b}(9390)$&$0^{-}$ &0,1,2,3&0&0& 9337 & 9931 & 10224 & 10471\\ \hline
$\Upsilon(9460)$&$1^{-}$ &0,1,2,3&0&1& 9460 & 10048& 10338& 10581 \\ \hline
$\chi_{b0}(9860)$&$0^{+}$&0,1,2,3&1&1& 9813 & 10110& 10359& 10591 \\ \hline
$\chi_{b1}(9893)$&$1^{+}$&0,1,2,3&1&1& 9931 & 10224& 10471& 10700 \\ \hline
$\chi_{b2}(9912)$&$2^{+}$&0,1,2,3&1&1&10048 & 10338& 10581& 10808 \\ \hline
$B_c(6277)$&$0^{-}$      &0,1,2,3&0&0& 6277 & 6719 & 6892 & 7025  \\ \hline
\end{tabular}
\end{table}
  
\begin{acknowledgments}

This work was supported by the DFG under Contract No. LY 114/2-1, 
by FONDECYT (Chile) under Grant No. 1140390 and  No. 1141280, 
by CONICYT (Chile) under Grant No. 7912010025, 
by CONICYT (Chile) Research Project No. 80140097 and 
by Tomsk State University Competitiveness Improvement Program. 
V.~E.~L. would like to thank Departamento de F\'\i sica y Centro
Cient\'\i fico Tecnol\'ogico de Valpara\'\i so (CCTVal), Universidad
T\'ecnica Federico Santa Mar\'\i a, Valpara\'\i so, Chile for warm
hospitality. 

\end{acknowledgments} 

\appendix\section{Details on kinematical variables: 
relations and $1/m_Q$ expansions} 

\subsection{Relation of $k^3$ and $x$} 
First we comment on the derivation of 
Eq.~(\ref{Jaus_eq}). It follows from a
formula relating $x$ and $k^3$ in the 
constituent quark rest frame~\cite{Jaus:1991cy}  
\eq 
x = \frac{1}{2} + \frac{k^3}{2E_k}\,.
\en 
Then, when substituting 
\eq
E_k = \sqrt{m_Q^2 + {\bf k}^2} 
= \sqrt{m_Q^2 + \bfk^2 + (k^3)^2} 
\en 
we get 
\eq
k^3 = (2x-1) E_k = (2x-1) \sqrt{m_Q^2 + \bfk^2 + (k^3)^2} 
\en 
and 
\eq 
k^3 = \frac{x-1/2}{\sqrt{x(1-x)}} \, \sqrt{\bfk^2+m_Q^2} \,. 
\en 

\subsection{Counting of momentum variables in the heavy quark mass expansion} 

Here we present the counting of momentum variables in the heavy 
quark mass expansion 
\eq\label{scaling} 
\hspace{-.6cm}
& &\bfk \sim {\cal O}(1) \,, \ k^3 \sim {\cal O}(1)\,, \ 
E_k = m_Q + {\cal O}(1/m_Q) \nonumber\\
\hspace{-.6cm}
& &x =  
\frac{1}{2} + \frac{k^3}{2E_k} = \frac{1}{2} + {\cal O}(1/m_Q) \,, 
\ x - \frac{1}{2} = {\cal O}(1/m_Q) \nonumber\\ 
\hspace{-.6cm}
& &\frac{1}{2} + \frac{k^3}{2E_k} = \frac{1}{2} + {\cal O}(1/m_Q) \,, 
\ \sqrt{x (1-x)} = \frac{1}{4} + {\cal O}(1/m_Q^2)\,, \nonumber\\
\hspace{-.6cm}
& &x + x' = 1 + \frac{k^3}{2E_k} + \frac{k^{'3}}{2E_{k'}}  
       = 1 + {\cal O}(1/m_Q) \,. 
\en 
Using the last formula in Eq.(\ref{scaling}) 
we introduce the delta-function $\delta(1-x-x')$ which imposes $1 = x+x'$ 
up to $1/m_Q$ corrections.

\end{document}